# Cavity optomechanical mass sensor in water with sub-femtogram resolution

Motoki Asano*, Hiroshi Yamaguchi, and Hajime Okamoto

[1]NTT Basic Research Laboratories, NTT Corporation, 3-1 Morinosato Wakamiya, Atsugi-shi, Kanagawa, 243-0198, Japan

E-mail: motoki.asano@ntt.com

Sub-femtogram resolution of an in-liquid cavity optomechanical mass sensor based on the twin-microbottle glass resonator is demonstrated. An evaluation of the frequency stability using an optomechanical phase-locked loop reveals that this cavity optomechanical sensor has the highest mass resolution of $(7.0 \pm 2.0) \times 10^{-16}$ g in water, which is four orders of magnitude better than that in our first-generation setup [Sci. Adv. 8, eabq2502 (2022)]. This highly sensitive mass sensor provides a free-access optomechanical probe in liquid and could thus be extended to a wide variety of *in-situ* chemical and biological metrology applications.

Cavity optomechanical devices allow high-precision optical detection of mechanical motion through photon pressure and/or the photothermal effect[1,2]. Measuring a mechanical frequency shift induced by an adhesion of small specimen on the devices enables highly sensitive mass detection, which can be utilized for various sensors[3-6]. One of the performance indices is given by the limit of detection, i.e., the minimum detectable mass, which is dependent on the frequency stability of the mechanical resonance. It is known that better performance is obtained in a device with a higher optical quality factor ($Q$), because the frequency stability is proportional to the signal-to-noise ratio (SNR)[7,8], which is enhanced in a high-$Q$ optical cavity.

For biological and chemical applications, the ability to operate such optomechanical sensors in liquid is crucial. However, it is difficult to straightforwardly extend this scheme to in-liquid conditions, because once the optical cavity is put into a liquid, the large refractive index and optical absorptance of liquids significantly reduce the optical $Q$. In pioneering work, this $Q$ reduction has been avoided by using a microdisk optical cavity made up of semiconductor materials with a high refractive index[9-11] or a hollow-core glass optical cavity in which a fluid channel does not directly couples to the optical cavity modes[12-14]. These approaches are based on a fixed-by-design architecture and thus suitable for passively detecting small specimens in liquid droplets[9-11] or in flowing liquid inside a channel[12-14], but they are not suitable for actively probing a specimen at a target location in liquid.

In contrast, a probe-type cavity optomechanical sensor in liquid has recently been demonstrated by utilizing a twin-microbottle resonator (TMBR) made of glass[15]. The TMBR has a unique structure, in which the mechanical resonances of the two microbottles are coupled with each other but their optical resonances are decoupled and isolated [Fig. 1(a)]. In the upper microbottle in contact with a tapered fiber, optically excited whispering gallery modes (WGMs) are coupled to mechanical radial breathing modes (RBMs) via photon pressure[15,16,17]. Therefore, by putting the lower microbottle into a liquid while the upper microbottle is kept in air, one can achieve highly sensitive optical detection of the RBM in the in-liquid microbottle through the mechanical coupling while avoiding the $Q$ reduction of the WGMs. This TMBR sensor allows us to actively access an arbitrary position in a liquid like a scanning probe microscope does[18,19], thus opening up a wide range of applications in mass sensing in liquid. However, the mass resolution of the previously reported TMBR is limited to the level of several picograms[15], which is not good enough to resolve small biological specimens, such as a single bacterium with a mass on the order of femtograms. Therefore, significant improvement of the mass resolution has been desired. In this letter, we report sub-femtogram resolution of the TMBR optomechanical mass sensor in water, which is four orders of magnitude better than that in the previously reported one. This is achieved by reducing the effective mass of the sensor, by introducing a balanced homodyne interferometer (BHI), and by optimizing the demodulation bandwidth of the lock-in amplifier used for the phase-locked loop (PLL).

The minimum detectable mass, i.e., the mass resolution $\delta m$, is given by $\delta m = 2m_{\text{eff}}\sigma_A$, where $m_{\text{eff}}$ is the effective mass of the mechanical mode, and $\sigma_A$ is the Allan deviation of the mechanical resonance frequency[7]. Therefore, the reduction of $m_{\text{eff}}$ directly contributes to improving mass resolution. Here, we fabricated a TMBR with the maximal diameter of 68 μm, which is about half the diameter of the previously reported TMBR [Fig. 1(b)]. To make this miniaturized TMBR, we first taper a commercial single-mode fiber with the clad diameter of 80 μm ($= D_c$) by the heat-and-pull process[15),16)] to a minimal diameter of 62 μm [Step 1, see Fig. 1(c)]. Next, we change the heat-and-pull position in such a way that one end of the taper overlaps with an end of a newly made second taper (Step 2). This step results in a microbottle structure with the maximal diameter ($D_b$) of 68 μm. We perform this step once again to make two microbottles sandwiched by three neck segments with a diameter $D_n$ of 62 μm (Step 3). The overall length of one of the fabricated microbottles is 400 μm, which is about a half that in the previously reported TMBR. Because the difference between $D_b$ and $D_n$ is sufficiently larger than the light wavelength of 1.5 μm, optical WGMs in the two microbottles are decoupled and isolated, while the mechanical RBMs in them are coupled with each other. The $m_{\text{eff}}$ of the RBMs is estimated from this TMBR geometry to be $4.0 \times 10^{-6}$ g, which is about an order of magnitude smaller than that in the previously reported TMBR[15].

Enhancing the frequency stability, i.e., reducing the frequency (Allan) deviation $\sigma_A$, is another approach to improving $\delta m$. Incorporating the BHI [Fig. 2(a)] is helpful for this because it enhances the SNR of the optical readout and thus contributes to reducing $\sigma_A$[7),8)]. We first compared the SNR of the thermal noise spectrum of the TMBR in air, measured with and without the BHI. An external cavity diode laser (ECDL) was used to probe the mechanical modes of the TMBR. This probe light with an optical power of about 1 mW was injected into a tapered fiber whose diameter had been downsized close to the light wavelength. The frequency of probe laser was set on the slope of the optical resonance whose $Q$ is $3.6 \times 10^6$ [Fig. 2(b)]. The output light from the tapered fiber was detected directly with an avalanche photodiode (APD) or through the BHI. In the case with the BHI, the non-resonant polarization component of the probe light plays the role of a local oscillator and is mixed with the signal light by adjusting the polarization with a polarizing beam splitter (PBS)[20]. Owing to the high-$Q$ optical resonance, thermal fluctuation of the two coupled mechanical modes was observed at around 56.7 MHz both with and without the BHI, but with a different SNR [Fig. 2(c)]. The two resonances have a similar amplitude and linewidth, where the theoretical fitting with the coupled mode model[15] shows that the eigenfrequency difference between the two microbottles (6.4 kHz) is much smaller than the linewidth (~140 kHz) and thus nearly tuned to form coupled mechanical resonators with the mechanical coupling strength of $g_M/2\pi = 225$ kHz. The optomechanical coupling rate, defined by $g_0 = \frac{\partial \omega_{opt}}{\partial x} x_{zpf}$ where $\omega_{opt}$ is the optical resonance frequency, and $x_{zpf}$ is the zero-point mechanical fluctuation, is estimated to be $g_0/2\pi = 0.4$ kHz with a calibration tone from a phase modulator (PM)[21]. Here, it is emphasized that the measurement with the BHI results in a better SNR than the measurement without it by a factor of five, leading to a noise floor level as low as $5 \times 10^{-20}$ m/$\sqrt{\text{Hz}}$ [Fig. 2(c)]. Hereafter, we used this sensitive BHI and evaluated the frequency stability of the TMBR when it was partially immersed into water. Before discussing the frequency stability, we show the properties of coupled RBMs with respect to the immersion depth in Fig. 3(a), where the origin of the immersion depth is set to the position where the bottom neck of the lower microbottle starts to make contact with the water surface. The change in the frequency and linewidth of the lower frequency mode is larger than that of the higher frequency mode [Figs. 3(b) and (c)] because of the eigenfrequency red-shift and viscous damping of the lower microbottle in water[15]. The observed change in frequency and linewidth is well consistent with the fluid-structure interaction theory for the incompressible liquid model [see the dashed curve in Fig. 3(b) and (c)][9),15].

The frequency stability in water was evaluated by keeping the immersion depth at 300 μm, where the frequency and linewidth of the lower frequency mode are respectively changed by -56 and +53 kHz from that in air [Figs. 3(b) and 3(c)]. In these measurements, an additional light was injected into the TMBR with the power of 10 mW to optically drive the mechanical modes. The optomechanical closed loop shown in Fig. 4(a) was also used to track the mechanical resonance frequency of the lower frequency mode and drive the RBM on resonance with the PLL. The optical intensity of this drive laser was modulated with an electro-optic intensity modulator (IM) while the probe light was detected with the BHI through an optical filter. The detected AC signal of the probe light through the BHI was sent to a lock-in amplifier (UHFLI, Zurich Instruments) and demodulated to extract the phase information with the PLL option. The rf signal output from the lock-in amplifier was sent back to the IM so that the driving frequency matched to the resonance of the lower frequency mechanical mode. During the measurement, the frequency of the probe laser was locked on the slope of the optical resonance through the DC signal via a servo controller. The Allan deviation was extracted from the averaged center frequency of the mechanical mode $f_0$ and the frequency deviation measured in time period $f(t+\tau) - f(t)$ as $\sigma_A(\tau) \equiv \sqrt{\langle (f(t+\tau) - f(t))^2 \rangle / 2f_0^2}$, where $\tau$ is the data integration time[22]. Once $\sigma_A$ is measured, $\delta m$ is estimated from the relation $\delta m = 2m_{\text{eff}}\sigma_A$.

Here, we show that $\sigma_A$ and $\delta m$ depend on the demodulation bandwidth $f_b$ of the lock-in amplifier used for the PLL[8),23)]. A characteristic time defined by the inverse of this bandwidth, $\tau_b = (\pi f_b)^{-1}$, gives the lower limit of the meaningful integration time ($\tau \geq \tau_b$) for evaluating the mass resolution that reflects the frequency deviation of the mechanical mode, whereas $\sigma_A$ at $\tau < \tau_b$ just shows the instability of the measurement setup[23]. Figure 4(b) shows the measured $\sigma_A$ and the corresponding $\delta m$ with respect to $\tau$ for three typical $\tau_b$ [64 ms ($f_b$ = 5 Hz), 0.64 ms ($f_b$ = 500 Hz), and 6.4 μs ($f_b$ = 50 kHz)]. The data were obtained when the voltage of the rf signal sent to the IM was set to 1.5 $V_{pp}$. The result clearly shows that the larger $\tau_b$ (i.e., the smaller $f_b$) causes the smaller $\sigma_A$ and $\delta m$. Note that $\sigma_A$ partially follows the line of $\sigma_A \propto \tau^{-1/2}$ [shown by dashed lines in Fig. 4(b)]. The $\tau^{-1/2}$ dependence indicates that the system is dominated by white noise[8),23)] and therefore $\sigma_A$ and $\delta m$ decrease with increasing $\tau$, which is due to the enhancement of the SNR with longer data integration. However, data integration time that is too long leads to non-negligible frequency drift and an increase in $\sigma_A$; thus, $\sigma_A$ and $\delta m$ have local minima at a certain $\tau$ above $\tau_b$. For $f_b$ = 5 Hz, the highest mass resolution of $\delta m = (7.0 \pm 2.0) \times 10^{-16}$ g is obtained at $\tau$ = 0.4 s [Fig. 4(b)]. This is about four orders of magnitude smaller than that for the previously reported TMBR[15], even smaller than the mass of a single bacterium. Note that the available minimum $f_b$ is 5 Hz (i.e., the maximum $\tau_b$ is 64 ms) and beyond that the phase is unlocked in the current PLL setup with the lock-in amplifier.

We also show that $\sigma_A$ and $\delta m$ depends on the driving power; the stronger the drive is, the smaller their values [Fig. 4(c)]. This is because the SNR increases with the stronger drive, as confirmed in the frequency response for three different rf voltages [Fig. 4(d)]. In the current PLL setup with the lock-in amplifier, the maximum rf drive voltage is limited to 1.5 $V_{pp}$. Therefore, the highest mass resolution is within the sub-femtogram level as described above. However, it can be easily expected that future improvements of the measurement setup will lead to better mass resolution with a modified SNR. Combining the setup with a TMBR whose size has been further reduced would allow unprecedented mass resolution close to the attogram level in liquid with this probe-type cavity optomechanical architecture.

In conclusion, we have demonstrated sub-femtogram mass resolution of an in-liquid cavity optomechanical sensor based on a twin-microbottle resonator made of glass. The mass resolution of $(7.0 \pm 2.0) \times 10^{-16}$ g, which is four orders magnitude smaller than previously reported[15], has been demonstrated with a parameter-optimized phase-locked loop in the miniaturized resonator. This probe-type cavity optomechanical mass sensor could be used for a wide range of applications of chemical and biological metrology in liquid.

**Acknowledgments**
The authors thank Koji Sakai, Riku Takahashi, and Aya Tanaka for fruitful discussions. This work was partly supported by JSPS KAKENHI (21H01023).

## References

1) M. Aspelmeyer, T. J. Kippenberg, F. Marquardt, Rev. Mod. Phys. 86, 1391 (2014).
2) I. Favero and K. Karrai, Nat. Photonics 3, 201 (2009).
3) F. Liu, S. Alaie, Z. C. Leseman, and M. Hossein-Zadeh, Opt. Express 21, 19555 (2013).
4) M. Sansa, M. Defoort, A. Brenac, M. Hermouet, L. Banniard, A. Fafin, M. Gely, C. Masselon, I. Favero, G. Jourdan, S. Hentz, Nat. Commun. 11, 3781 (2020).
5) D. Navarro-Urrios, E. Kang, P. Xiao, M. F. Colombano, G. Arregui, B. Graczykowski, N. E. Capuj, M. Sledzinska, C. M. Sotomayor-Torres, G. Fytas, Sci. Rep. 11, 7829 (2021).
6) W. Yu, W. C. Jiang, Q. Lin, T. Lu, Nat. Commun. 7, 12311 (2016).
7) M. Sansa, E. Sage, E. C. Bullard, M. Gély, T. Alava, E. Colinet, A. K. Naik, L. G. Villianueva, L. Duraffourg, M. L. Roukes, G. Jourdan, S. Hentz, Nat. Nanotech. 11, 552 (2016).
8) P. Sadeghi, A. Demir, L. G. Villanueva, H. Kähler, and S. Schmid, Phys. Rev. B 102, 214106 (2020).
9) E. Gil-Santos, C. Baker, D. T. Nguyen, W. Hease, C. Gomez, A. Lemaître, S. Ducci, G. Leo, I. Favero, Nat. Nanotech. 10, 810–816 (2015).
10) E. Gil-Santos, J. J. Ruz, O. Malvar, I. Favero, A. Lemaître, P. M. Kosaka, S. García-López, M. Calleja, J. Tamayo, Nat. Nanotech. 15, 469 (2020).
11) S. Sbarra, L. Waquier, S. Suffit, A. Lemaître, I. Favero, Nano Lett. 22, 710 (2022).
12) K. H. Kim, G. Bahl, W. Lee, J. Liu, M. Tomes, X. Fan, T. Carmon, Light Sci. Appl. 2, e110 (2013).
13) K. Han, K. Zhu, G. Bahl, Appl. Phys. Lett. 105, 014103 (2014).
14) J. Suh, K. Han, G. Bahl, Appl. Phys. Lett. 112, 071106 (2018).
15) M. Asano, H. Yamaguchi, and H. Okamoto, Sci. Adv. 8, eabq2502 (2022).
16) M. Asano, Y. Takeuchi, W. Chen, Ş. K. Özdemir, R. Ikuta, N. Imoto, L. Yang, T. Yamamoto, Laser Photon. Rev. 10, 603 (2016).
17) A. J. R. MacDonald, B. D. Hauer, X. Rojas, P. H. Kim, G. G. Popowich, J. P. Davis, Phys. Rev. A 93, 013836 (2016).
18) P. K. Hansma, J. P. Cleveland, M. Radmacher, D. A. Walters, P. E. Hillner, M. Bezanilla, M. Fritz, D. Vie, H. G. Hansma, C. B. Prater, J. Massie, L. Fukunaga, J. Gurley, V. Elings, Appl. Phys. Lett. 64, 1738 (1994).
19) J. K. H. Horber, M. J. Miles, Science 302, 1002 (2003).
20) A Schliesser, G. Anetsberger, R Rivière, O. Arcizet, and T J Kippenberg, New. J. Phys. 10, 095015 (2008).

21) M. L. Gorodetksy, A. Schliesser, G. Anetsberger, S. Deleglise, T. J. Kippenberg, Opt. Express 18, 23236 (2010).

22) D. W. Allan, Proc. IEEE 54, 221 (1966).

23) A. Demir and M. S. Hanay, IEEE Sens. J. 20, 1947 (2020).

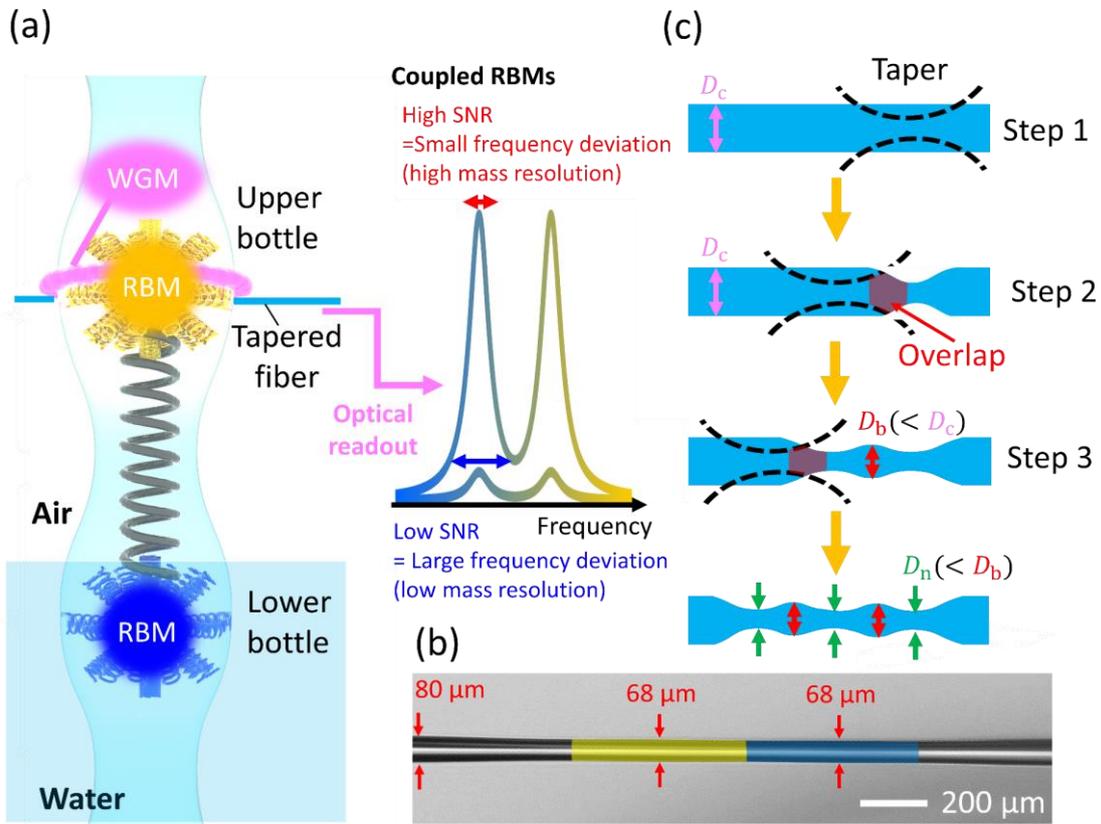

**Fig. 1.** (a) Conceptual illustration of the TMBR partially immersed in water. The mechanical RBMs are mutually coupled via the center neck of the TMBR. The coupled mechanical modes are optically detected via radiation pressure on the upper bottle by reading out the laser light using a tapered fiber contacting the TMBR. (b) Optical microscope image of the TMBR. The blue and yellow shaded areas correspond to each microbottle structure. (c) Fabrication procedure for the miniaturized TMBR using the heat-and-pull method. Overlapping the ends of two tapered areas results in a microbottle structure with the maximum diameter, $D_b$, smaller than the initial diameter of the silica fiber, $D_c$. Continuing this process leads to the two interconnected microbottles through the center neck with the diameter, $D_n$, smaller than $D_b$.

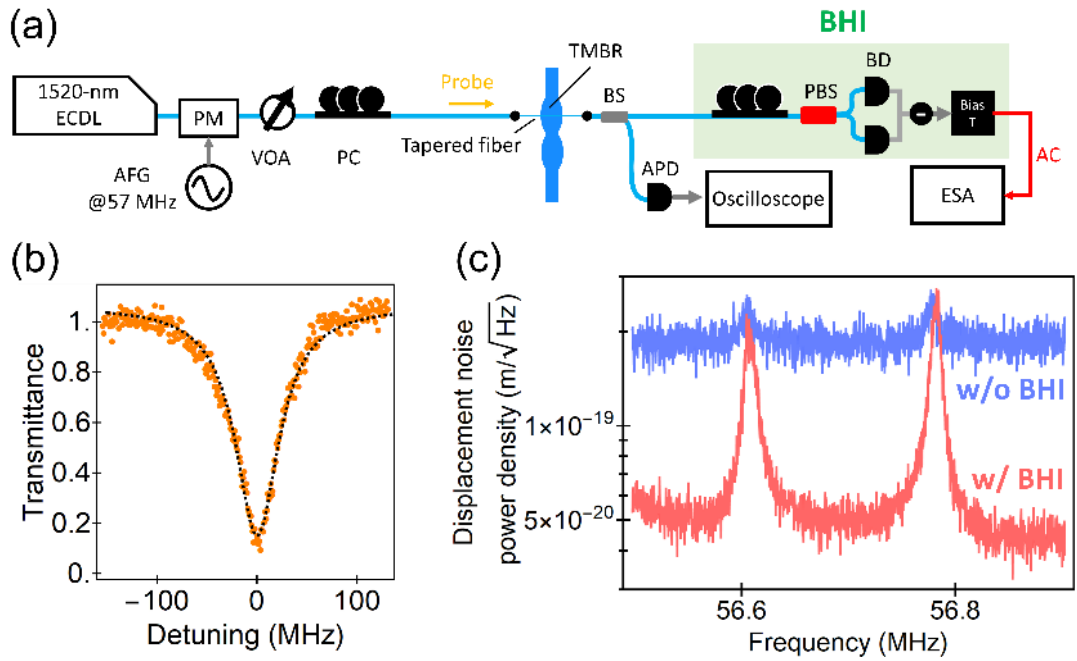

**Fig. 2** (a) Schematic of the setup for measuring the thermal displacement noise of the TMBR in air with a balanced homodyne interferometer (BHI). ECDL: external cavity died laser. PM: phase modulator. AFG: arbitrary function generator. VOA: variable optical attenuator. PC: polarization controller. BS: beam splitter. APD: avalanche photodiode. PBS: polarizing beam splitter. BD: balanced detector. ESA: electric spectrum analyzer. (b) Transmission spectrum of an optical WGM. (c) Thermal displacement noise power spectrum of the coupled RBWs measured with the BHI (red) and without it (blue).

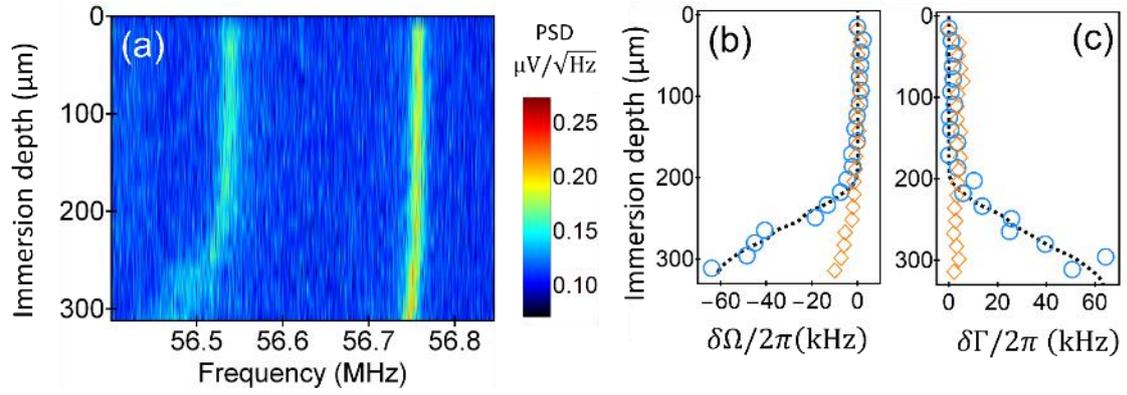

**Fig. 3** (a) Immersion depth dependence of thermal displacement noise power spectrum of the coupled RBWs. (b), (c) Frequency shift and linewidth broadening with respect to the immersion depth. The blue circles and orange squares show the values for the lower and higher frequency modes, respectively. The dashed black curves are the theoretical fitting with the fluid-structure interaction model.

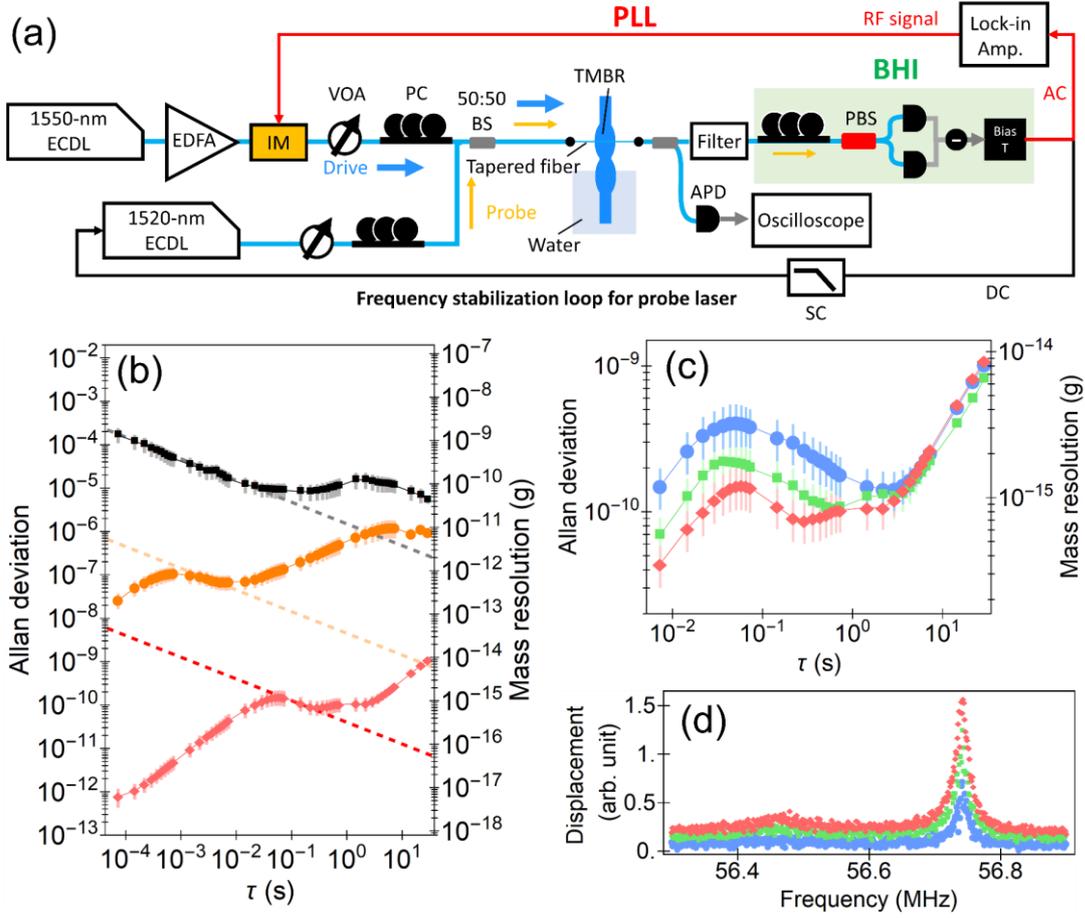

**Fig. 4** (a) Schematic of the setup for evaluating frequency stability with the optomechanical PLL. EDFA: erbium dipole fiber amplifier. IM: intensity modulator. SC: servo controller. (b) Allan deviation and the corresponding mass resolution with respect to the data integration time, τ, for three different demodulation bandwidths (black squares, 50 kHz; orange circles, 500 Hz; red diamonds, 5 Hz) with the drive voltage of 1.5 $V_{pp}$. The error bars show the standard deviation of Allan deviation in each τ. The dashed lines show the $\tau^{-1/2}$ dependence. (c) Allan deviation and the corresponding mass resolution with respect to τ for three different drive voltages (blue circles, 0.5 $V_{pp}$; green squares, 1.0 $V_{pp}$; red diamonds, 1.5 $V_{pp}$) with the measurement bandwidth of 5 Hz. The error bars show the standard deviation of Allan deviation in each τ. (d) Frequency response of the coupled RBMs for three different drive voltages (blue circle, 0.5 $V_{pp}$; green squares, 1.0 $V_{pp}$; red diamonds, 1.5 $V_{pp}$).